\documentclass[10pt,conference]{IEEEtran}
\IEEEoverridecommandlockouts
\usepackage{algorithmic,algorithm}
\usepackage{cite}
\usepackage[utf8x]{inputenc} 
\usepackage{amsmath}
\usepackage{amssymb,amsmath}
\usepackage{graphicx}
\usepackage{epsfig}
\usepackage{grffile}
\usepackage{balance,color}
\usepackage{mathtools, nccmath}
\DeclarePairedDelimiter{\nint}\lfloor\rceil

\providecommand{\keywords}[1]{{\textit{Index Terms}}}
\abovecaptionskip 
\belowcaptionskip 

\begin{document}

\title{I-Health: Leveraging Edge Computing and Blockchain for Epidemic Management} 
\author{Alaa Awad Abdellatif$^{*\dag}$, Lutfi Samara$^{\ddag}$, Amr Mohamed$^{*}$, Aiman Erbad$^{*}$, Carla Fabiana Chiasserini$^{\dag}$,\\  Mohsen Guizani$^{*}$, Mark Dennis O'Connor$^{+}$, and James Laughton$^{+}$ \\ 
$^*$Department of Computer Science and Engineering, Qatar University \\
$^\dag$Department of Electronics and Telecommunications, Politecnico di Torino  \\
$^\ddag$Department of Electrical Engineering, Qatar University \\
$^+$Mobile Healthcare Service, Hamad Medical Corporation, Qatar \\
\thanks{ 
This work was made possible by NPRP grant \# NPRP12S-0305-190231 from the Qatar National Research Fund (a member of Qatar Foundation). The work of Mark Dennis O'Connor and James Laughton was supported by Abhath Project \# MRC 01-17-091 from Hamad Medical Corporation. The findings achieved herein are solely the responsibility of the authors. } 
} 
\maketitle

\begin{abstract}
Epidemic situations typically demand intensive data collection and management from different locations/entities within a strict time constraint. 
Such demand can be fulfilled by leveraging the intensive and easy deployment of the Internet of Things (IoT) devices.  The management and containment of such situations also rely on cross-organizational and national collaboration.  
Thus, this paper proposes an Intelligent-Health (I-Health) system that aims to aggregate diverse e-health entities in a unique national healthcare system by enabling swift, secure exchange and storage of medical data. In particular, we design an automated patients monitoring scheme, at the edge, which enables the prompt discovery, remote monitoring, and fast emergency response for critical medical events, such as emerging epidemics. Furthermore, we develop a blockchain optimization model that aims to optimize  medical data sharing between different health entities to provide effective and secure health services. Finally, we show the effectiveness of our system, in adapting to different critical events, while highlighting the benefits of the proposed I-Health system. 

\end{abstract}
\begin{IEEEkeywords}
Blockchain, edge computing, Internet of Things (IoT), priority assignment, outbreak management.  
\end{IEEEkeywords}

\section{Introduction\label{sec:Introduction}} 

Advances in e-health and Internet of Things (IoT) technologies can play an integral, crucial, and evolving role in providing swift responses to outbreaks and health crises. In light of the recent pandemic, the development of smart, efficient and secure health system for the purpose of managing and stopping the spread of such crises becomes a worldwide interest.  
A pivotal contribution towards the development of intelligent health system can be achieved by automating most of the healthcare functions to provide efficient healthcare services. Emerging technologies, such as Artificial Intelligence (AI), Edge Computing, and Blockchain, can turn this vision into reality. Such technologies can transform the traditional health system into an Intelligent-Health (I-Health) system that enables effective collection, management, and sharing of medical data during outbreaks. Indeed, I-Health can support diverse functions, including event detection and characterization, real-time remote monitoring, as well as identification and management of patients with high mortality risks.    

In the era of I-Health, all health-related services should be managed in efficient and distributed ways. Specifically, during the periods of epidemics, an intensive amount of data will need to be gathered (from diverse IoT devices), analyzed, and shared across multiple entities to conduct in-depth medical studies, epidemic investigation, and improving the response time in emergency conditions. Moreover, such systems are of extreme importance since it is critical to monitor the patients' status precisely outside medical centers to minimize the patients' visits, and hence minimizing the risks of physical contact with the patient. 
Thus, we envision that improving the communication links between patients and healthcare providers is mandatory to enable large-scale healthcare services and personalized medicine.  
However, remote accessibility of medical data and Electronic Health Records (EHRs) by different entities comes with processing, communications, and security challenges. Typically, traditional healthcare systems implement weak security measures which jeopardizes the security of the overall system. For instance, from 2016 to 2017, the number of reported health-related attacks increased by 89\% as reported in ‎\cite{ransomware}. 

In this work, we argue that designing an efficient, secure, and decentralized I-Health system fulfilling the aforementioned challenges can be implemented by leveraging edge computing and blockchain technologies. We envision that bringing the intelligence close to the users/patients, using edge computing, along with sharing the important data over a blockchain network is a key for detecting and managing urgent outbreaks\footnote{The term outbreak in this work refers to any acute public health problem requiring urgent epidemiological investigation, including infectious disease outbreaks such as the Coronavirus Outbreak (COVID-19) \cite{WHO}.}.
On one hand, blockchain is a decentralized ledger of transactions that are shared among multiple entities while preserving the integrity and consistency of the data through smart contracts ‎\cite{WorldEconomicForum}. Hence, it effectively supports data processing and storage at different entities as well as their interconnections. Blockchain also provides traceability and audibility of transactions from multiple organizations, which plays a crucial role in tracking the supply chain of certain drugs/vaccine during adverse events. 
On the other hand, being decentralized allows for the potential application of edge computing, which enables a swift and portable emergency detection through identifying and monitoring infected individuals at the edge.  

We therefore aim at paving the way to design an efficient I-Health system that addresses the above aspects through:
\begin{enumerate}
	\item Designing a secure and decentralized I-Health system that relies on blockchain and edge computing technologies to provide early detection, fast response, and intelligent management for urgent epidemic situations. 
	\item Developing an automated patients monitoring scheme at the edge. The proposed scheme allows for an accurate detection of the changes in the patients' records, hence ensures a fast notification about the patient’s state, at the edge-level, while sharing important information with the different participating entities in the system.  
	\item Developing a multi-channel blockchain architecture with a flexible, optimized configuration model, which allows for:  
	(i)   assigning different priorities for the acquired transactions based on their urgency level and importance; 
	(ii)  optimizing blockchain channels configuration to adapt to diverse types of applications/data with different characteristics.  
	\item Demonstrating the effectiveness of the proposed system in improving the performance of healthcare systems using a real-world dataset.    
\end{enumerate} 

In the rest of the paper, we begin by introducing the main challenges that will be tackled in this paper, then introducing our I-Health architecture and framework (Section \ref{sec:Sec2}).   
 Then, Section \ref{sec:system} presents our patients monitoring scheme, while Section \ref{sec:Configuration} introduces our blockchain optimization model with the priority assignment task. Performance evaluation of our system is then discussed in Section \ref{sec:sim_results}. The related work and benefits of the proposed I-Health scheme are presented in Section \ref{sec:Related}. Finally, the paper is concluded in Section \ref{sec:conclusion}.  


\section{I-Health Challenges, Architecture, and Framework \label{sec:Sec2}}

In this section, we first highlight the key challenges of managing infectious disease epidemics, then we present our I-Health architecture and  framework to address these challenges.    
 
\subsection{Challenges of emerging infectious disease epidemics \label{sec:why}} 

To track and control the spread of an epidemic (e.g., dangerous infectious diseases), piles of information from diverse locations (e.g., hospitals, clinics, and airports) as well as reports concerning disease outbreaks should be collected, processed, and analyzed. 
However, acquiring and sharing such amount of information between different e-health entities at different geographical locations is challenging due to: data quality, availability, timeliness, and completeness. 
Moreover, for effective epidemic management, an e-health system must: (i) expedite the process of information collection and investigation; (ii) provide a fast response with high quality service level and security for the entire population.    
To this end, the following issues have to be adequately addressed using the proposed I-Health system.

\textbf{Limited resources:}  
During the times of the spread of infectious diseases (such as the recent COVID-19 outbreak \cite{WHO}), most of the hospitals are required to serve hundreds of patients daily. This could generate an intense load on the hospitals for a long time. Furthermore, such outbreaks that can spread from human to human can put the medical staff at high risk of being infected. In some recent outbreaks \cite{pandamic2007}, a number of healthcare facilities were shut down to prevent their staff from contracting the virus, rendering the traditional healthcare systems futile in such critical times.         
   
\textbf{Secure connectivity:}   
  During an epidemic, secure communications is a critical tool to detect and handle the virus spreading as early as possible \cite{Cybersecurity2020}. Indeed, real-time access to a patient's EHRs enables e-health systems to give timely care to the patients through the nearest point of care.   
However, medical data exchange across multiple organizations imposes major challenges on the system design in terms of network load and security. Thus, innovative methods for secure data access, analysis, and management are needed to handle the enormous amounts of data from different locations, which also help the medical staff to focus on epidemiological investigation.     

\textbf{Monitoring infected patients:} 
One major aspect for managing the spread of epidemics is the precise monitoring of infected patients that are part of the epidemic investigation. Hence, healthcare systems must support efficient monitoring for the patients' state, in a timely manner, even outside the hospitals.    


\begin{figure*}[htp]
	\centering
		\scalebox{3.3}{\includegraphics[width=0.27 \textwidth]{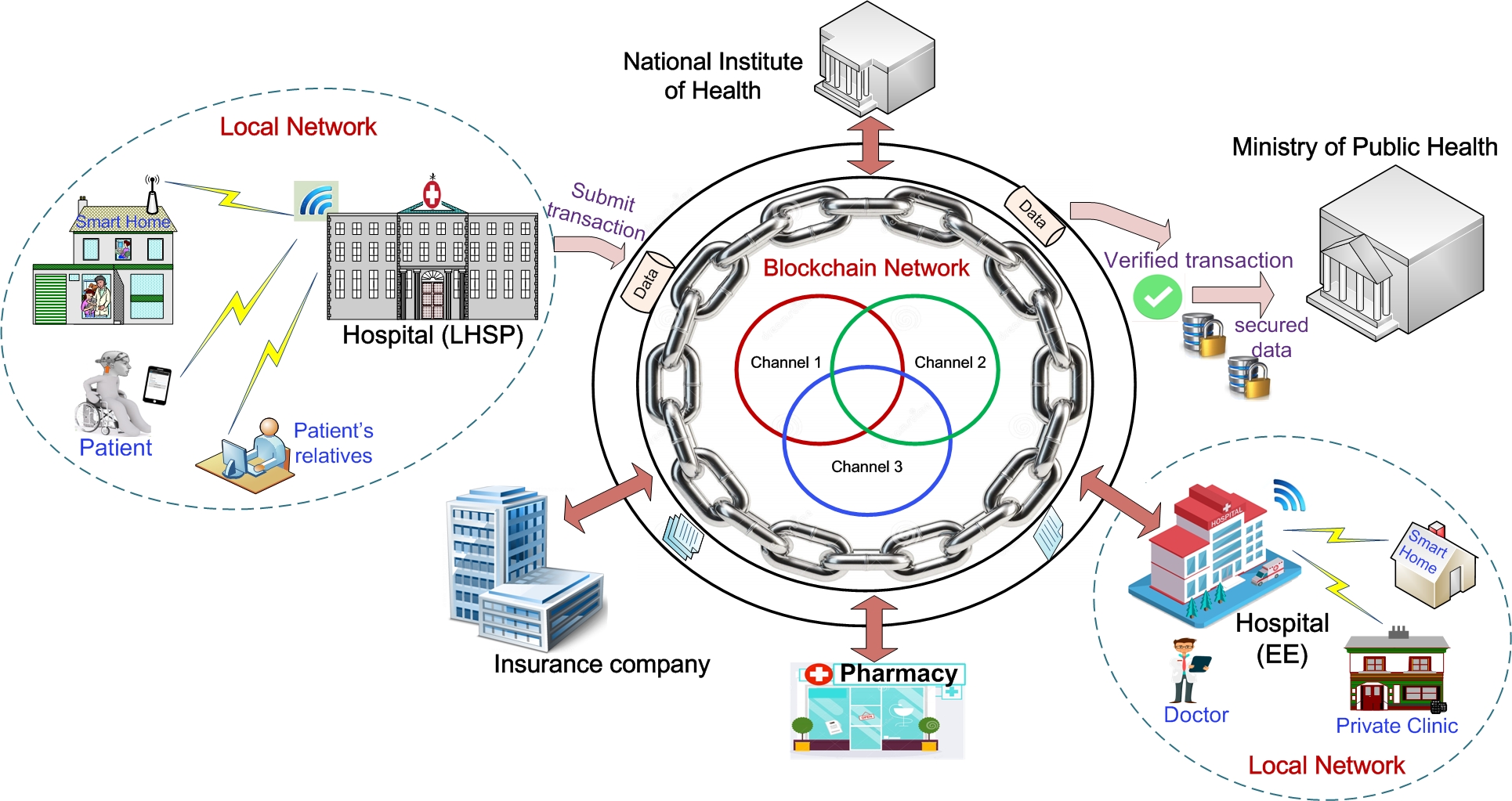}}
	\caption{The proposed I-Health system architecture. }
	\label{fig:system_model}
\end{figure*}

\subsection{I-Health architecture\label{sec:Architecture}} 

To address the above challenges, we propose the following I-Health architecture, which is comprised of diverse e-health entities whose fundamental role is to monitor, promote, and maintain people's health. 
The proposed I-Health system architecture, shown in Figure~\ref{fig:system_model}, is divided to two main networks: (a) a Local network, and (b) a blockchain network.  
For the sake of scalability, we consider that the intended e-health entities gather health-related data from the local network, process these data, and share important information through the blockchain network. The shared data are validated and stored locally by the various entities in the blockchain, which are trusted entities with large storage and computational capabilities \cite{suggested}.

The local network stretches from the data sources located on or around  patients to the Local Healthcare Service Provider (LHSP), like e.g., a hospital. It contains the following major components:  \\ 
\textit{a.1) Internet of Medical Things (IoMT):} A combination of IoT devices attached/near to the patients to be leveraged for monitoring health conditions and activities within the smart assisted environment. Examples include: body area sensor networks (i.e., implantable or wearable sensors that measure different biosignals and vital signs),  smartphones, IP cameras, and external medical and non-medical devices. \\       
\textit{a.2) Local Healthcare Service Provider (LHSP):} 
An LHSP is a medical facility which monitors and provides the required healthcare services for the local patients, records the patients' state, and provides prompt emergency services if needed. Most importantly, the LHSP plays a significant role in monitoring the patients' state not only inside the medical facility (intra-medical-facility patient care), but also outside such facilities, as e.g. home patient care related services. Also, it can be connected with the private clinics that may transfer patients to it for more advanced care, or even with the patient's close circle to follow up on the patient's conditions. 

As far as the blockchain network is concerned (see Figure~\ref{fig:system_model}), the core is the multi-channel blockchain-based data sharing architecture that enables secure access, processing, and sharing of medical data among diverse e-health entities.  
Blockchain is indeed  particularly suitable for secure medical data sharing because of its immutability and decentralization features, which are perfectly consistent with our proposed I-Health architecture. Using blockchain, all transaction blocks (i.e., containing health-related information) can be securely shared, accessed, and stored by physicians, decision makers, and other healthcare entities. The latter include, but are not limited to:   \\   
\textit{b.1) External Edge (EE):}   
In the proposed architecture, a hospital or a LHSP have more advanced tasks than the ones mentioned above: it can act as an EE that is responsible for data storage, applying sophisticated data analysis techniques, and sharing important health-related information with public health entities. Hence, leveraging the power of edge computing, each entity can verify the authenticity and integrity of the medical data at the EE before sharing it within the blockchain. \\      
\textit{b.2) Ministry of Public Health (MOPH):} 
The main role of MOPH is monitoring the quality and effectiveness of healthcare services through coordination with different health entities. MOPH waives the responsibility of healthcare services to the hands of public and private health sectors while regulating, monitoring, and evaluating their healthcare services to guarantee an acceptable quality of care. \\   
\textit{b.3) Insurance companies:} 
 One important aspect for e-health systems is integrating healthcare providers, patients, and payers into one ``digitized community" in order to improve the quality of services and drive down the costs. Indeed, to realize a sustainable healthcare-business model, healthcare providers will have to own health plans powered by insurance companies. \\
\textit{b.4) Other entities:}
Different entities can be also part of our I-Health system, such as National Institutes of Health (NIH) and pharmacies. The former are major players in clinical research and health education, while the latter have to coordinate with prescribers and/or private insurance companies to confirm the dosage and formulation (e.g., liquid or tablet), or to submit insurance claims and ensure payment.   
{   


\subsection{The proposed I-Health framework \label{sec:Edge}} 

 The ultimate goal of our I-Health system is to fulfill diverse challenges of epidemics mentioned above through implementing the following main functionality at the edge and blockchain (see Figure~\ref{fig:Framework}):  
(i) data collection, feature extraction, and patients' state monitoring, in order to ensure high-reliability and fast response time in emergency detection;    
(ii) secure data accessibility anytime and anywhere to different entities.

We envision that integrating edge computing with blockchain in our I-Health framework provides a potential solution to all of the aforementioned challenges. Indeed, leveraging edge computing allows for defining when and what data to share through the I-Health system. This is essential for ensuring that the most important and up-to-date information is available for investigation.   
In this context, we propose an automated patients' state monitoring scheme at the edge, which enables:
\begin{enumerate}
	\item collecting the data of different patients (inside or outside the hospital);
	\item identifying specific features from the acquired data that are informative and pertinent to the patients' state;
	\item detecting major changes in the patients' state leveraging the identified features. 
\end{enumerate}
   
After processing the acquired information, at the edge, we define the critical events that should be shared with other entities through permissioned blockchain. A general blockchain architecture mainly consists of: data sender, Blockchain Manager (BM), and validators. First, the data senders upload their data, in a form of ``transactions'', to the nearby BM. Then, the BM acts as a validators' manager: it distributes unverified blocks to the validators for verification, triggers the consensus process among the validators, and inserts the verified block in the blockchain \cite{AB3}. Hence, the BM acts as the leader, while the validators are the followers that cooperate to complete the block verification task.  
In our framework, we consider a multi-channel blockchain, where each channel corresponds to a separate chain of transactions that can be used for enabling data access and private communications among the channel users \cite{Multichannels}. Leveraging such architecture allows for treating different health-related events effectively. 
In particular, we consider three channels in our blockchain, channel 1 for urgent data (such as emergency notifications), channel 2 for non-urgent data but requiring a high security level (such as confidential legal messages), and channel 3 for normal data. 
Accordingly, we propose three new tasks at the BM: 
\begin{enumerate}
	\item priority assignment, which aims to assign different priority levels for the received transactions from diverse entities based on their urgency level and arriving time;
	\item blockchain channel allocation, which allocates the received transactions to the appropriate channel based on their urgency and security levels;
	\item blockchain configuration optimization, where different blockchain configuration parameters are optimized based on diverse application requirements and data types.    
\end{enumerate}
We remark that the BM has a logical role that any entity in the proposed architecture can take on, possibly by taking turns, or that can be taken by the leading organization that wants to share its data \cite{Incentivizing2019}. 

In what follows, we present how the above functionality can be implemented at the edge and BM.  

\begin{figure}[t!]
	\centering
		\scalebox{1.83}{\includegraphics[width=0.27 \textwidth]{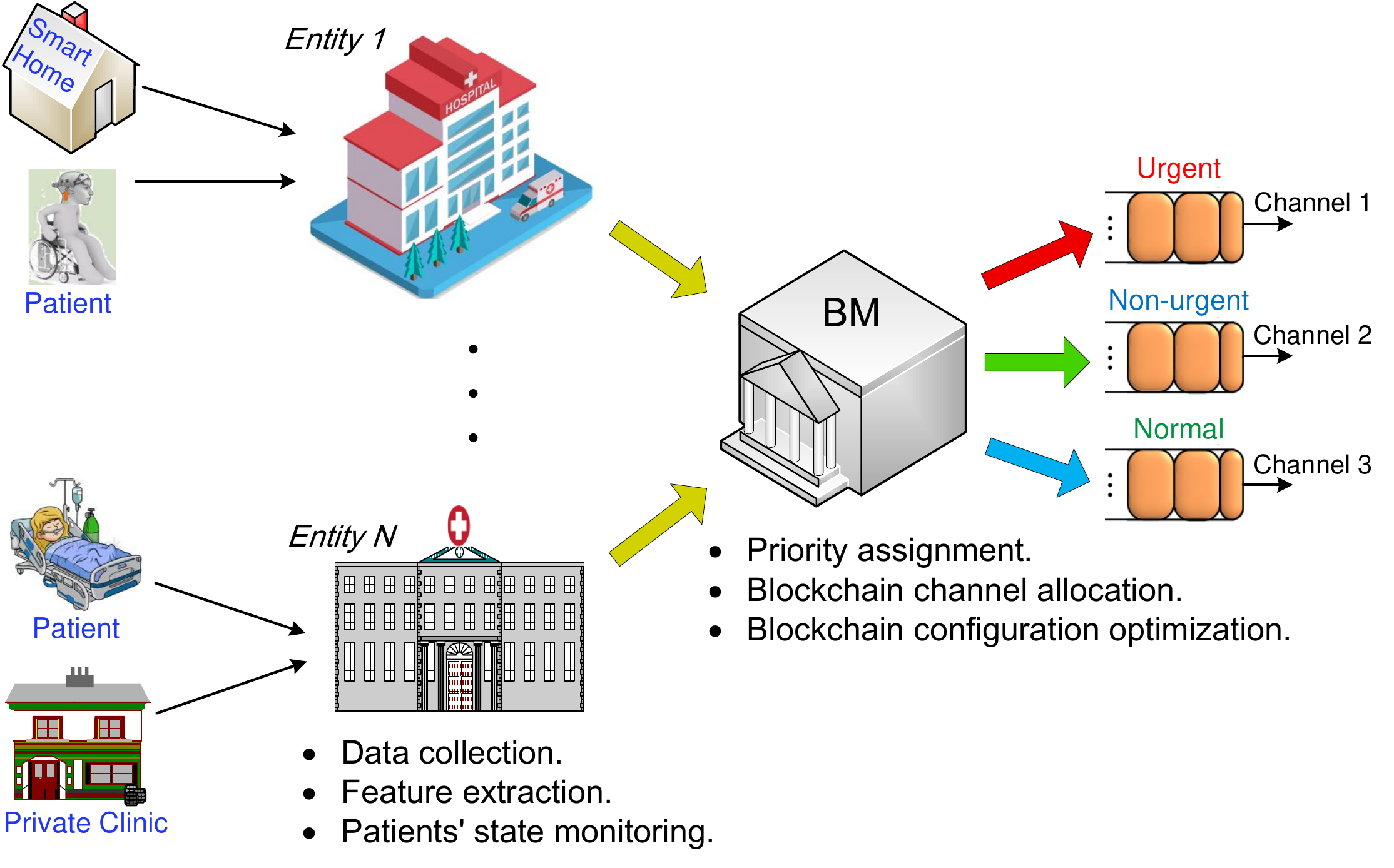}}
	\caption{Diagram representing the proposed I-Health framework, highlighting the different tasks performed by the edge and BM, as well as the corresponding data flow.}
	\label{fig:Framework}
\end{figure}

\section{Implementing The Edge Functions \label{sec:system}}

This section presents the first stage in our framework, which focuses on the edge functionality. 
We consider a specific case study related to remote monitoring. During epidemics, it is crucial to move large number of patients with mild symptoms into home care. If I-health system can adequately monitor this large number of patients, from different locations, it will conserve hospitals' facilities to absorb critical cases, which may help save more lives during outbreaks.            
Thus, we propose an efficient, low-complexity and automated patients monitoring scheme at the edge. The proposed scheme defines a change indicator, which measures the percentage of change in patient's records from one period to the next. 

Our scheme has been designed leveraging biological data that has been collected from patients undergoing routine planned treatment. The acquired data includes 14-channel Electroencephalography (EEG) signals and routine observational data, such as temperature, blood pressure, and so on. Monitoring EEG signals provides an additional source of information to help in detecting changes of the patients' state, and to monitor the dosage of hypnotic drugs \cite{EEGbook}.   
Our data has been collected from 30 patients taking a specific medication during three different sessions. The three sessions represent the data of a patient \textit{before}, \textit{during}, and \textit{after} taking the medication. More description about the data collection is presented in Section \ref{sec:sim_results}. However, without loss of generality, the proposed scheme and methodology can be easily applied to different types of data.  
The proposed scheme comprises the following main steps.       

\subsection{Feature extraction\label{sec:Feature}}  
The first step in our changes detection scheme is identifying the main statistical features that are informative, representative, and pertinent to EEG changes detection.  
As shown by the signal behavior in Figure \ref{fig:With_time}, it is difficult for the doctors to differentiate and detect the changes. However, after analyzing these signals, we found that they exhibit different mean, variance, and amplitude variations. Moreover, it is crucial to consider as relevant features the Root Mean Square (RMS), i.e., a good signal strength estimator, and kurtosis, i.e., a measure of the tailedness of the probability distribution.  
We therefore select the following four features, in addition to the minimum $x_{ij}^{min}$ and maximum $x_{ij}^{max}$ values of the acquired data:  \\ 
\noindent
\textsl{Mean} 
\begin{equation}
\mathcal{M}_{ij} = \frac{1}{\mathfrak{N}} {\sum_{k=1}^{\mathfrak{N}} x_{ij}(k)},
\label{mean}
\end{equation}
\textsl{Variance}
\begin{equation}
\sigma_{ij}^2 = \frac{1}{\mathfrak{N}} {\sum_{k=1}^{\mathfrak{N}} \left|x_{ij}(k)-\mathcal{M}_{ij}\right|^2}, 
\label{Variance}
\end{equation} 
\textsl{Root mean square}
\begin{equation}
R_{ij} = \sqrt{\frac{1}{\mathfrak{N}}\sum_{k=1}^{\mathfrak{N}} {\left|x_{ij}(k)\right|}^2},  
\end{equation}
\textsl{Kurtosis}
\begin{equation}
\nu_{ij} =  \frac{ \frac{1}{\mathfrak{N}} \sum_{k=1}^{\mathfrak{N}} \left( x_{ij}(k)-\mathcal{M}_{ij} \right)^4 }{ \left( \frac{1}{\mathfrak{N}} {\sum_{k=1}^{\mathfrak{N}} (x_{ij}(k)-\mathcal{M}_{ij})^2} \right)^2 },  
\label{Kurtosis}
\end{equation}
where $x_{ij}(k)$ is the values of input EEG signal for channel $i$ and patient $j$, and $\mathfrak{N}$ is the number of samples. Accordingly, for a given patient $j$, the above  features will be calculated, for each EEG channel $i$, to represent the patient's state over a time window of $N$ samples.   

\begin{figure}[t!]
	\centering
		\scalebox{1.6}{\includegraphics[width=0.27 \textwidth]{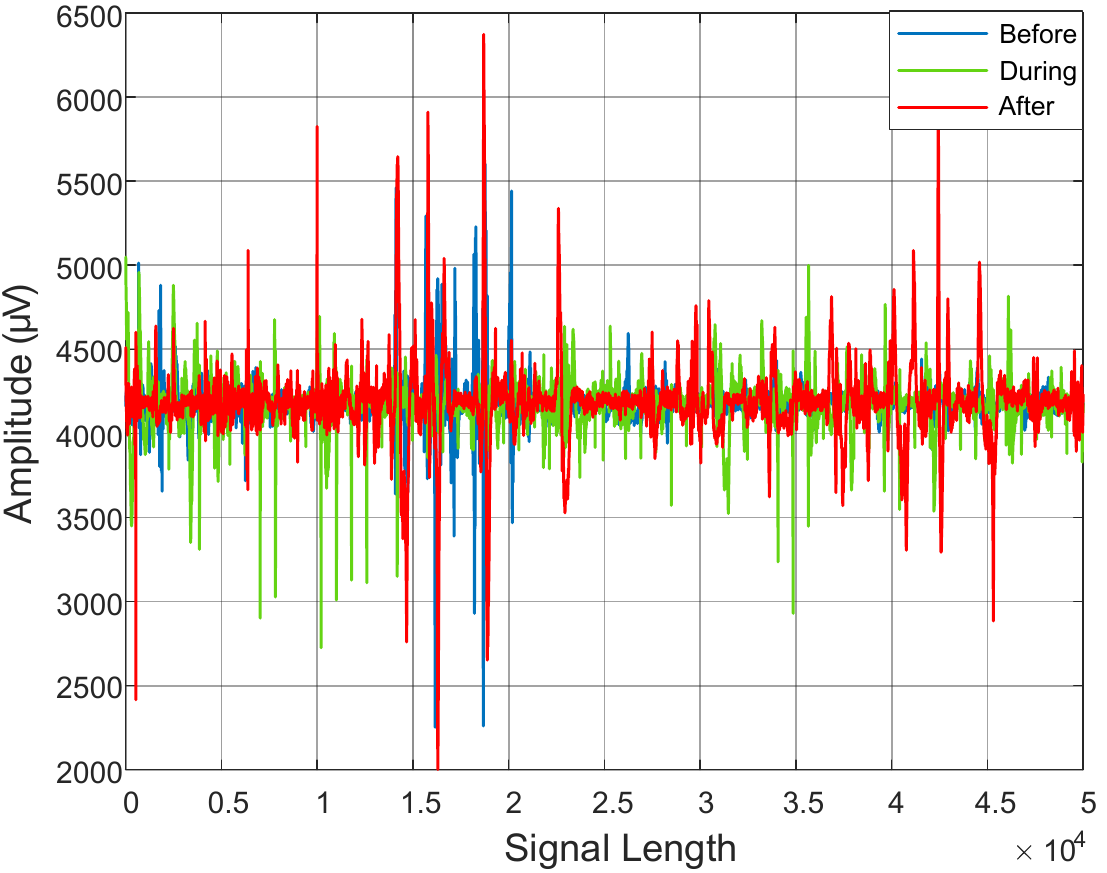}}
	\caption{An example of the acquired EEG signals, from one channel, in time domain: before, during, and after given the medication to a patient. }
	\label{fig:With_time}
\end{figure}

\subsection{Changes detection and sharing\label{sec:changes}}  
The second step in our scheme is detecting, at the edge, the major changes in the patient's state. Hence, based on the detected changes, the edge node (i.e., a hospital) can optimize what to share on the blochchain, as follows:
\begin{itemize}
	\item in case of detecting major changes (i.e., of an emergency), it will share through blockchain an emergency notification, along with the raw data that may require further investigation;
	\item in case of detecting minor/no changes, it will share only the obtained features;  
  \item in case of detecting major changes in one or two channels only, it means that the measurements may be inaccurate due to some errors in the experiment. Thus, it is recommended to notify the responsible physician to repeat the measurements.  
\end{itemize}

We exploit the extracted features to perform an initial detection to the major changes in EEG signals at the edge. The advantages of our scheme is two-fold. First, by detecting the changes in the acquired data at the edge, we can significantly decrease the amount of information to be shared on the blockchain, without missing important information in case of emergency. Second, in case of emergency, a quick alert and notification can be initiated based our scheme, hence facilitating effective analysis without wasting the physician's time. 

The fundamental question now is: How can we obtain a simple yet accurate classification rule using the generated features to reveal the major changes in the acquired data? 
First, we define a statistical indicator $\delta_{ij}$, for an EEG channel $i$ and patient $j$, that integrates generated features as follows: 
\begin{equation}
\delta_{ij} = {\mathcal{M}_{ij} + \sigma_{ij}^2+ R_{ij}+\nu_{ij}+x_{ij}^{min}+x_{ij}^{max}}. 
\label{eq:delta} 
\end{equation}
Using (\ref{eq:delta}), we define a change indicator vector $\mathcal{K}_j=\left[\kappa_{1j} \cdots \kappa_{Cj} \right]$ for a patient $j$, where 
$\kappa_{ij}$ is defined as
\begin{equation}
\kappa_{ij} = \left[ \frac{\left|\delta_{ij}^b - \delta_{ij}^d\right|}{\bar{\delta}}  + \frac{\left|\delta_{ij}^d - \delta_{ij}^a\right|}{\bar{\delta}}\right]\times 100, 
\label{eq:kappa} 
\end{equation}
where 
\begin{equation}
\bar{\delta}= \frac{\sum_{j=1}^{P}\sum_{i=1}^{\mathcal{C}} \delta_{ij}^b+\delta_{ij}^d+\delta_{ij}^a}{3\mathcal{C}P}. 
\end{equation}
In (\ref{eq:kappa}), $\bar{\delta}$ is the statistical mean of $\delta$, acquired during offline training, for all channels $i \in \left\{1, \cdots, \mathcal{C} \right\}$ over all patients $j \in \left\{1, \cdots, P \right\}$.  

Second, we define a classification rule using the obtained $\mathcal{K}_j$ to detect the major changes/errors of the acquired EEG data, where $\mathcal{K}_j$  will represent  the  condition  part  of the rule, while the status of the patient $\omega_j$ will represent its consequent part. Accordingly, we obtain through our experiments the following classification rule
\begin{equation}
\omega_j = \begin{cases}
     \text{Major},&  \text{if } ||[\mathcal{K}_j - \zeta]^+||_0 > 2  \\  
		 \text{Minor},&  \text{if }  ||[\mathcal{K}_j - \zeta]^+||_0 = 0 \\ 
     \text{Repeat},&  \text{if } 0 < ||[\mathcal{K}_j - \zeta]^+||_0 \leq 2,
\end{cases}
\label{class_rule}
\end{equation}
where $[\mathbf{a}]^+ = \max(0,\mathbf{a})$ provides a vector of either positive or 0 elements in a vector $\mathbf{a}$, $||.||_0$ is the zero$^{th}$ norm operator, and $\zeta$ is a threshold that assesses the major changes in the EEG signal (e.g., we consider $\zeta=30\%$).  

 We remark that this scheme will be exploited to obtain the status of the patient at the edge, hence optimizing what to share through blockchain. Moreover, it provides a quick detection for the major changes in the patient's state, while keeping the complexity low, hence it is amenable for implementation at any mobile edge. 


\section{Blockchain Optimization: Priority assignment and Solution \label{sec:Configuration}}  

The second stage in our framework is developing an optimized blockchain configuration model that enables sharing of different health-related events and information among diverse healthcare entities.    
We envision  that  for designing an efficient I-Health system, the acquired data from various entities should be treated in different ways, based on their urgency and security levels. 
For example, urgent data (i.e., require minimum latency) should be given highest priority and dealt with a restricted blockchain, i.e., with minimum number of validators. On the contrary, for low priority types of data but requiring a high security level, fully restricted blockchain should be used (see Figure~\ref{fig:Edgetasks}). In case of normal data, i.e., that has requirements on both latency and security, an optimized blockchain configuration is used.     
We remark that data types and emergency levels are defined at the edge by applying different data classification, event detection, and summarization techniques, as shown in Section \ref{sec:Edge}.  
In general, the more validators participate in the block verification stage, the higher the security level is, but also the larger the latency (due to the verification delay) and the higher the cost (due to verification fees) that are experienced \cite{BC2019}, \cite{moreverifier}. Instead, as the number of transactions per block grows, the latency increases, while the  cost per transaction decreases \cite{moreverifier}. 
Accordingly, the proposed blockchain optimization addresses the aforementioned challenges by designing an event-driven secure data sharing scheme, as detailed below.

The proposed scheme draws on the BM concept \cite{BC2019}, which acts as a validators' manager, that is responsible for:   
\begin{enumerate}
	\item gathering the transactions from different entities,
	\item assigning different priorities to the gathered transactions based on their urgency level,   
	\item updating the blockchain configuration considering urgency and security level of the gathered transactions,
	\item preparing and distributing unverified blocks to the selected validators (e.g., hospitals, NIH, and MOPH, which have sufficient computation and storage resources), 
	\item interacting with the validators to complete block verification tasks.
\end{enumerate}
  Thus, the BM is a critical component in our scheme, which dynamically updates the blockchain configuration's parameters, based on the diverse applications' requirements and data types, such that the optimal trade-off among security, latency, and cost is obtained. 
Also, we remark that, in line with the traditional consensus scheme, the validators take turns in working as BM for a given time period \cite{BC2019}.  
		
\begin{figure}
	\centering
		\scalebox{1.86}{\includegraphics[width=0.27 \textwidth]{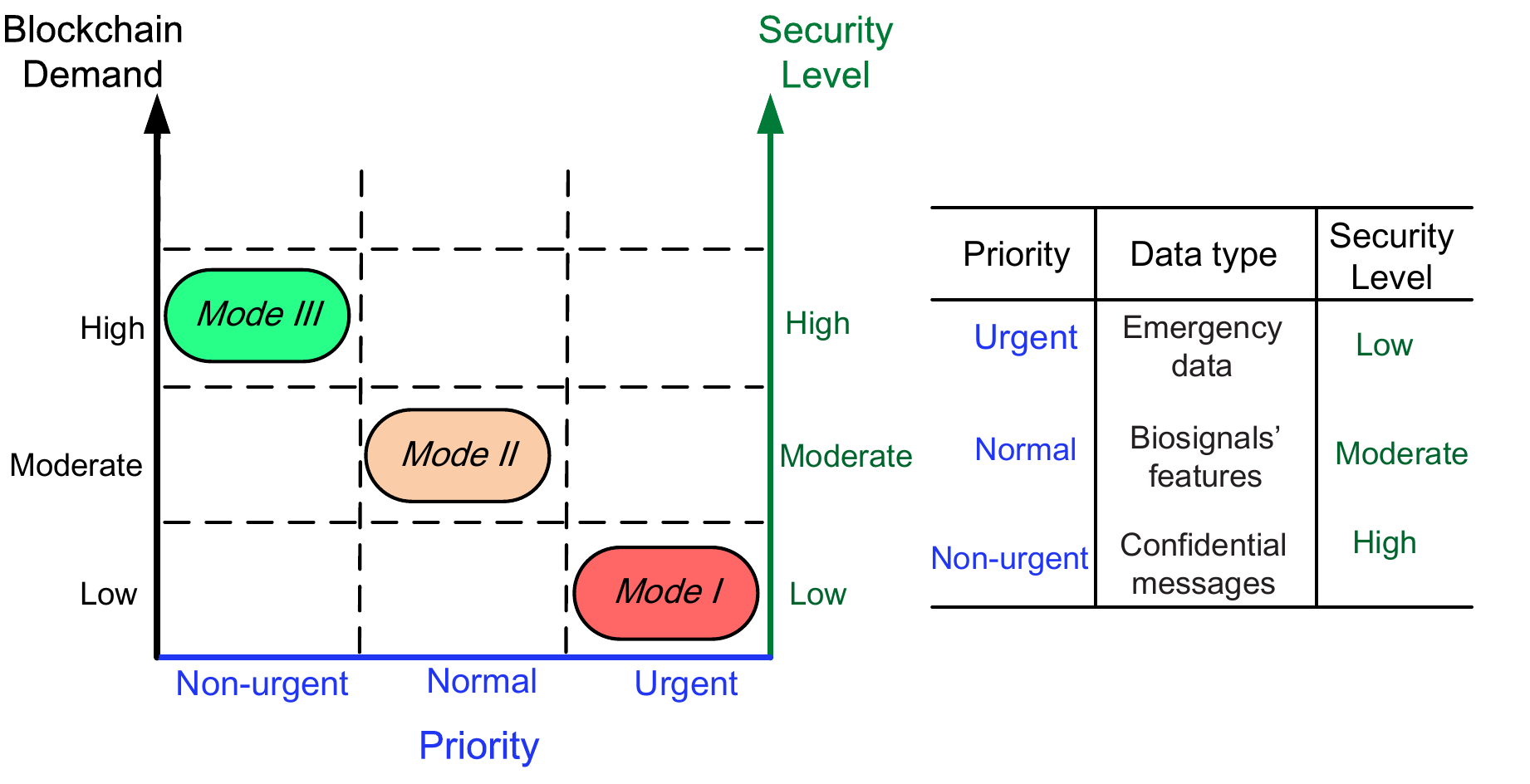}}
	\caption{Blockchain modes based on the data priority and required security level.}
	\label{fig:Edgetasks}
\end{figure}

\subsection{Priority assignment} 

Before optimizing the blockchain configuration's parameters, we highlight the role of priority assignment task at the BM. This task  aims to minimize the sojourn time of the received transactions from different entities based on their urgency level. Herein, the sojourn time refers to the total amount of time a transaction is expected to wait before being added to the blockchain.  
This sojourn time will be controlled by identifying different urgency levels, namely urgent, normal and non-urgent. Then, we adopt the use of queuing models to calculate the sojourn time based on the urgency levels of different received transactions. 
In particular, we define the sojourn time based on the preemptive-resume priority concept \cite{adan2002queueing}, i.e., the transactions with a higher priority interrupts the processing of transactions with lower priorities.  

It is assumed that $N$ entities (e.g., hospitals) are sending their transactions to the BM, each with an arrival rate $\lambda_i$, for $i \in \left\{1, \cdots, N\right\}$.  
All received transactions from different entities are temporarily stored in the BM's buffers. In this paper, buffer overflows are negligible since it is assumed that $\sum_{i=1}^{N} {\lambda_i} < \mu$, where $\mu$ is the service rate at the BM.  
By adopting the well-established M/M/1 queuing model \cite{malandrino2019reducing} (and the references therein) for the received transactions with equal priorities, the average sojourn time of entity $i$ is defined as 
\begin{align}
{S}_i^{e} = \frac{1}{\mu - \sum_{i=1}^{N} {\lambda_i}}. 
\end{align}
However, to handle the received transactions efficiently, the BM assigns different priorities for them based on their urgency levels and corresponding entity weight\footnote{Entity weight can represent the degree of influence that an entity has on the national health system}. Hence, transactions with high urgency and coming from high impact entities will be assigned the highest priority.  
%
To derive the average sojourn time for transactions with different priorities, we start from the general expression of the sojourn time which we denote by ${S}_i^{g}$, that can be calculated by applying \cite[{Sec. 9.2}]{adan2002queueing}
\begin{align}\label{eq:sp0} \nonumber
{S}_i^{g} &= \frac{   \sum_{n=1}^{i} \lambda_i {R}_i  }{ (1-(\frac{\lambda_1}{\mu} + \hdots + \frac{\lambda_i}{\mu})) (1-(\frac{\lambda_1}{\mu} + \hdots + \frac{\lambda_{i-1}}{\mu}))   }\\ 
& + \frac{{B}_i}{1-(\frac{\lambda_1}{\mu} + \hdots + \frac{\lambda_{i-1}}{\mu})}, 
\end{align}
where ${R}_i$ and ${B}_i$ are the mean service and mean residual service times of the $i^{th}$ entity, respectively.  
The adopted M/M/1 queuing model implies that we have exponential service times with mean ${B}_i=1/\mu$ and ${R}_i=1/\mu$ \cite{adan2002queueing}. Hence, substituting the aforementioned results in (\ref{eq:sp0}) yields the following average sojourn time expression
\begin{align}\label{eq:sp} \nonumber
{S}_i &= \frac{  \frac{1}{\mu} \sum_{n=1}^{i} \lambda_i  }{ (1-(\frac{\lambda_1}{\mu} + \hdots + \frac{\lambda_i}{\mu})) (1-(\frac{\lambda_1}{\mu} + \hdots + \frac{\lambda_{i-1}}{\mu}))   }\\ 
& + \frac{\frac{1}{\mu}}{1-(\frac{\lambda_1}{\mu} + \hdots + \frac{\lambda_{i-1}}{\mu})}.
\end{align}

To assess the benefits of the proposed urgency priority assignment compared to equal priority assignment, we present Figure~\ref{fig:sojourn}, which depicts the average sojourn time versus the entity ID. 
In this figure, we simulate the arrival rate of 21 different entities, where each entity is assigned a different priority based on its urgency level. In particular, it is assumed that entities 1 through 8 $\in$ urgent, entities 9 through 12 $\in$ normal, and entities 13 through 21 $\in$ non-urgent. Moreover, the packet arrival rate per entity is assumed to be a constant that equals to 2 transactions/s.  
The obtained results show that unlike the equal priority assignment, which obtains the same sojourn time for all entities, the proposed urgency priority assignment yields a significant reduction in sojourn time, especially for entities with an ``urgent'' status. We also observe that for the transactions belonging to low priority entities, the sojourn time is increased, when compared to that of the equal priority, which makes sense since it is tagged with low urgency (non-urgent). The figure also shows the effect of varying the average service rate on the obtained sojourn time. It is clear that the sojourn time increases when the service rate decreases, however, using our urgency priority assignment allows for decreasing the sojourn time of most of the entities (only three entities will have higher sojourn times than that of the equal priority assignment). 

We remark that service rate $\mu= n/L$, where $n$ is the number of transaction per block, and $L$ is the block verification latency inside the blockchain. Thus, optimizing blockchain configuration will have direct impact on the obtained sojourn time, as will be shown later.    
  
\begin{figure}
	\centering
		\scalebox{1.7}{\includegraphics[width=0.27 \textwidth]{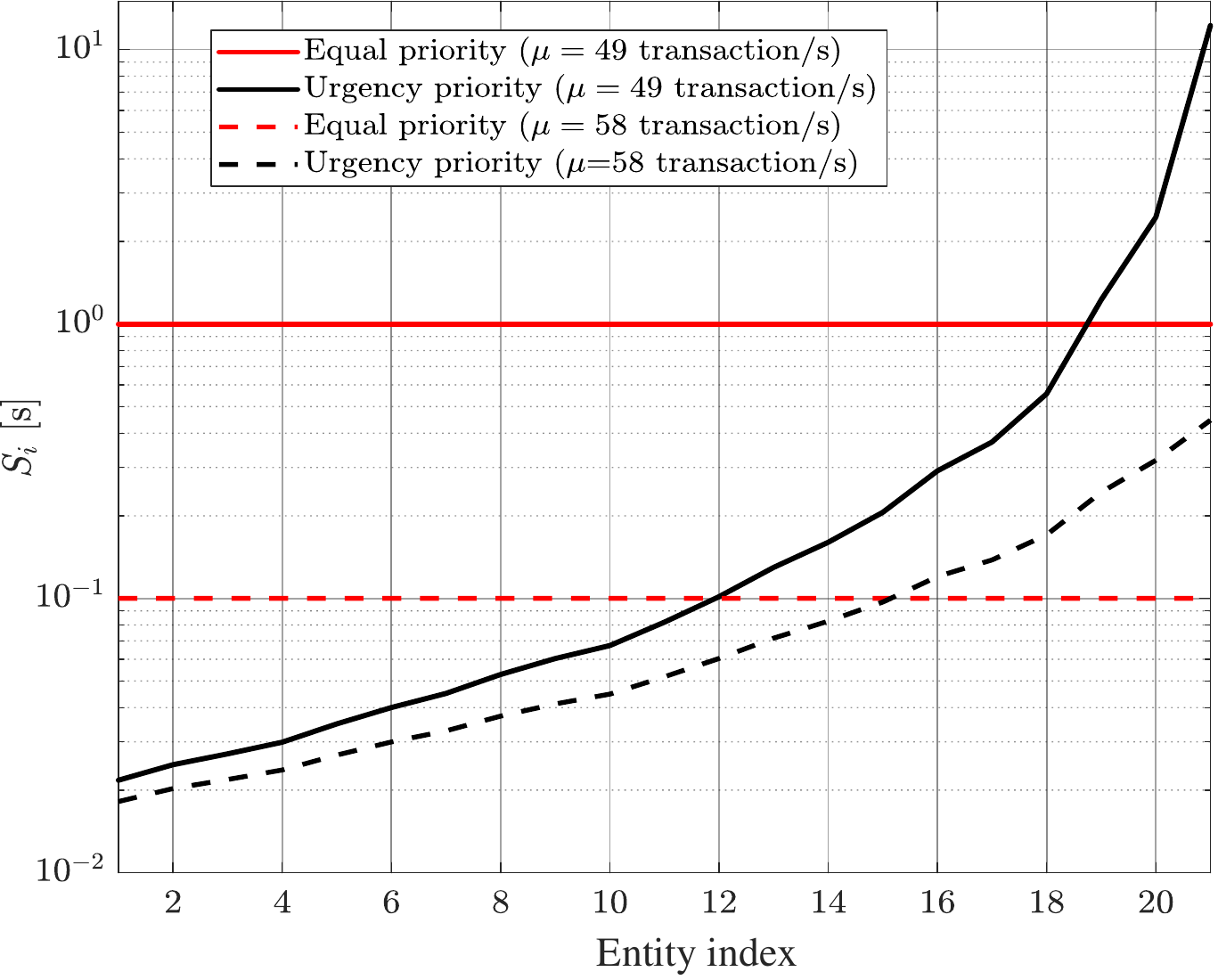}}
	\caption{The obtained average sojourn time for different entities using equal priority and urgency priority assignments, while varying service rate $\mu$.}
	\label{fig:sojourn}
\end{figure}
\subsection{Optimal blockchain configuration  \label{sec:Solution}}  

Given the received transactions with different priorities, the BM aims at mapping these transactions into different configurations of the blockchain.     
The proposed blockchain optimization model considers permissioned blockchain with Delegated Proof-of Stake (DPoS) consensus algorithm\footnote{Consensus algorithm is a process of ensuring the integrity and consistency of the blockchain across all participating entities \cite{AB3}.}, which performs the consensus process using pre-selected validators \cite{BC2019}. 
Our model focuses on three main metrics at the BM, namely,  latency ($L$), security ($\eta$), and cost ($C$). However, these metrics have different values and units, which must be first normalized with respect to their maximum values (denoted by $l_m$, $\eta_m$, and $c_m$, respectively) to make them comparable. 
Then, to deal with such conflicting metrics, we define an aggregate utility $U$, which combines them into a single function:
\begin{equation}
U = \alpha \cdot \frac{L}{l_m} + \beta \cdot \frac{\eta_m}{\eta} + \gamma \cdot \frac{C}{c_m},  
\label{eq:utility}
\end{equation} 
where $\alpha$, $\beta$, and $\gamma$ are weighting parameters representing the relative importance of the considered metrics, such that $\alpha+\beta+\gamma=1$. Also, $m$ is the number of selected validators, with  maximum and minimum values equal to $M$ and $v$, respectively, and $n$ is the number of transactions per block, with  maximum and minimum values equal to $\chi$ and $t$, respectively. 
Accordingly, the BM can obtain the best blockchain configuration, by solving the following optimization problem:
\begin{eqnarray}
\mbox{\bf P:} &&    \min_{m, n} \left( U \right) \label{eq:obj}
\hfill \\ 
\mbox{s.t.} &&	c_i \geq \rho_i \cdot x_i,  \ \  \forall i \in \left\{1, \cdots, m \right\} \label{eq:C1} \hfill \\  
&&	v \leq m \leq M,  \label{eq:C2}  \hfill \\  
&& t \leq n \leq \chi.   \label{eq:C3}  \hfill
\end{eqnarray}
In (\ref{eq:obj}), the cost function is defined as  
$  C = \frac{\sum_{i=1}^{m} c_i}{n}$,  where $c_i$ is the computational cost of validator $i$ to finish the verification task, while the security level is defined as 
$ \eta = {\theta \cdot m^q}$,  
where $\theta$ is a coefficient given by the system, and $q\geq 2$ is an indicator factor representing the network scale.   
$L$ refers to the latency of the block verification process, which includes: (i) unverified block transmission from the BM to validators, (ii) block verification time, (iii) verification result broadcasting and comparison between validators, and (iv) verification feedback transmission from the validators to BM \cite{BC2019}. Hence, the latency is defined as  
\begin{equation}
L = { \frac{n \cdot B}{r_d}+ \max_{i \in \left\{v, \cdots, M\right\}} \left({\frac{K}{x_i}}\right) + \psi(n \cdot B)m + \frac{O}{r_u} },
\label{eq:latency}
\end{equation}
where $B$ is the transaction size, $K$ is the required computational resources for block verification task, $x_i$ is the available computational resources at validator $i$, ${O}$ is the verification feedback size, $r_d$ and $r_u$ are, respectively, the downlink and uplink transmission rates from the BM to the validators and vice versa. In (\ref{eq:latency}), $\psi$ is a predefined parameter that can be obtained using the statistics belonging to the previous processes of block verification (as detailed in \cite{BC2019}).  
Finally, in our architecture, it is assumed that the validators are offloading their computational load of the verification process to the cloud/fog providers (CFPs). Hence, validator $i$ should buy the required computing resources $x_i$ from a CFP in order to access these resources from the remote cloud or the nearby fog computing unit \cite{Cloudprice2019}. Thus, for validator $i$ to participate in the verification process, it should receive a cost $c_i$ that at least covers its payment to the CFP. This condition is represented in constraint (\ref{eq:C1}), where $\rho_i$ represents the payment from validator $i$ to the CFP, in order to acquire the needed resources for the verification process.  
     
According to the acquired data types and application's requirements, the weighting coefficients $\alpha$, $\beta$, and $\gamma$ are defined. Hence, the optimal number of validators $m^*$ and transactions per block $n^*$ can be obtained by solving the proposed optimization problem. However, the above optimization problem  is an integer programming optimization, which is an NP-complete problem \cite{Integeroptimiz}.    
In light of the problem complexity, we propose below a light-weight iterative approach for obtaining an efficient solution of the formulated problem.  
 
In order to efficiently solve the formulated problem in (\ref{eq:obj}), we look at the problem as a block size optimization, as a function of $n$, and a block verification optimization, as a function of $m$. The block verification variable can be considered as a global variable that is relevant to the overall blockchain process, while the block size variable is a local variable at the block preparation phase. We therefore decompose the problem into the block size and block verification sub-problems, such that each of them is a function of one decision variable only and, hence, can be solved independently of the other. Then, an efficient-iterative algorithm is proposed for obtaining the optimal solution of (\ref{eq:obj}) by leveraging the proposed problem decomposition.   

Starting by the block size problem, a closed-form expression for the solution can be obtained by imposing that the derivative with respect to $n$ of the objective function is equal to 0, while considering $m$ as a constant. I.e., 
\begin{eqnarray} 
\partial{}/\partial{n} \left[ \alpha \cdot {L} + \beta \cdot \eta^{-1} + \gamma \cdot C \right ] & = & 0 \nonumber\\
\alpha\left(\frac{B}{r_d}+\psi \cdot B \cdot m\right) - \gamma \frac{\sum_{i=1}^{m} \rho_i \cdot x_i }{n^2} & = & 0 \nonumber\\
\frac{\gamma {\sum_{i=1}^{m} \rho_i \cdot x_i}}{\alpha(\frac{B}{r_d}+\psi \cdot B \cdot m)} & = & n^2.  
\label{derivative}
\end{eqnarray}
Thus, the optimal $n$ is given by:
\begin{equation}
	n = \sqrt{\frac{\gamma {\sum_{i=1}^{m} \rho_i \cdot x_i}}{\alpha(\frac{B}{r_d}+\psi \cdot B \cdot m)}}.
	\label{eq:n}
\end{equation}

Considering block verification optimization, an efficient Blockchain Configuration Optimization (BCO) algorithm is proposed (see Algorithm \ref{alg:BCO}). BCO algorithm leverages the idea of problem decomposition to find the optimal solution of (\ref{eq:obj}) in practical scenarios, where different validators have different verification response time. 
The main steps of BCO algorithm can be summarized as follows:
\begin{enumerate}
	\item BM distributes unverified blocks to the validators. 
	\item Validators that finish block verification faster are selected one by one.
 \item Given the selected validators ($m$), ${n}$ is calculated, using (\ref{eq:n}), and approximated to the nearest integer. Then, $n^*$ is obtained, such that the constraint in (\ref{eq:C3}) is satisfied. 
\item After adding a new validator, we check the ``\textit{gain}" condition, i.e., the obtained reduction in the security term (i.e., $\beta \cdot \eta^{-1}$) is greater than the obtained increase in the latency and cost terms (resulting from adding the new validator). If the ``\textit{gain}" condition is satisfied, this validator is added to the selected validators, otherwise it is discarded and $m^*$ is obtained.    
\end{enumerate}
We remark that the maximum number of iterations for the BCO algorithm to converge to the optimal solution is $M$, thanks to the derived closed-form solution for $n^*$.  

\begin{algorithm}
\caption{Blockchain Configuration Optimization (BCO) algorithm}
\label{alg:BCO}
\begin{algorithmic}[1]
\STATE {\textbf{Input:} $x_i$, $\rho_i$, $v$, $M$, $t$, $\chi$.}
\FOR {$m = v+1:M$}
\STATE {Calculate ${n}$ using (\ref{eq:n}).}
\IF {  $\nint{n} < t$.}
\STATE {$n^*=t$.}
\ELSIF {  $\nint{n} > \chi$.}
\STATE {$n^*= \chi$.}
\ELSE 
\STATE {$n^*= \nint{n}$.}
\ENDIF
\IF {$\beta \cdot \eta^{-1}(m-1) - \beta \cdot \eta^{-1}(m) < (\alpha \cdot {L}(m) + \gamma \cdot C(m))-(\alpha \cdot {L}(m-1) + \gamma\cdot C(m-1))$}
\STATE {$m^*=m-1$.}
\STATE Break \% $m^*$ is obtained
\ENDIF
\ENDFOR
\STATE {\textbf{Output:} $m^*$, $n^*$.}
\end{algorithmic}
\end{algorithm}

}

\begin{figure*}
	\centering
		\scalebox{3.7}{\includegraphics[width=0.27 \textwidth]{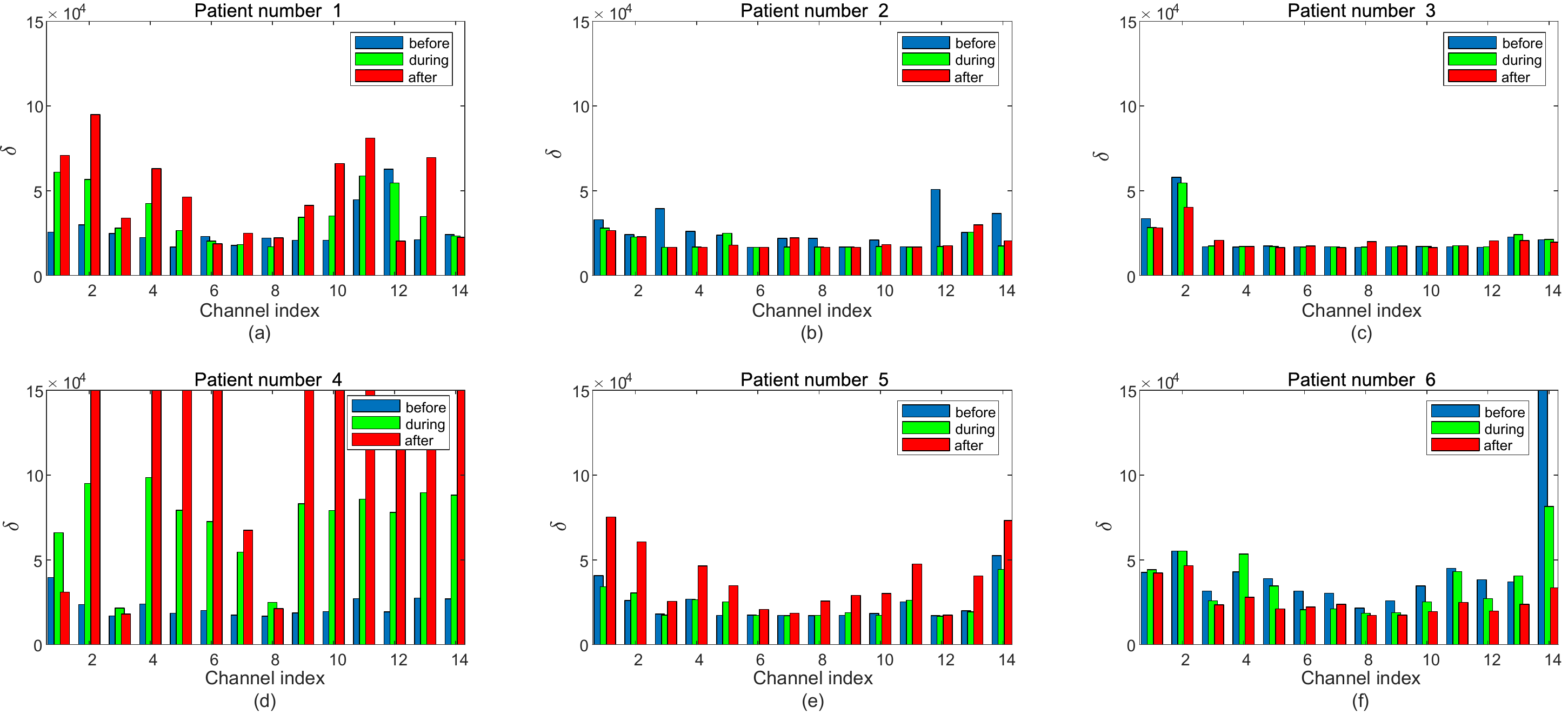}}
	\caption{The variations of change indicator $\delta$, over different channels, for six patients: before, during, and after taking a medication. }
	\label{fig:CI}
\end{figure*}

\section{Simulation Results \label{sec:sim_results}} 

For our performance evaluation, we use the data in \cite{ourdata} that has been collected from patients undergoing routine planned treatment. The data collection process has been carried out in the patient recovery center of Hamad Medical Corporation \cite{HMC}.   
 The  acquired data has been collected using EMOTIV EPOC+, which comprises 14 EEG channels (i.e., electrodes)\footnote{EEG monitoring are conducted using EEG electrodes, which gather and record the electrical activity of the brain. The acquired EEG signals are amplified, digitized, and then sent to a computer for processing and storage \cite{EPOC}.} for whole brain sensing \cite{EPOC}, in addition to the routine observational data such as temperature and blood pressure. This data has been collected from 30 patients receiving intravenous antibiotic medication. Each patient has been monitored for 30 minutes: before, during, and after taking the medication.  
Moreover, our results were generated considering 21 entities, where the packet arrival rate per entity is assumed to be uniformly distributed with mean equals to 1 transactions/s. Other simulation parameters are set as follows: $q=4$, $O=0.5$ Mb, $\theta=1$, $r_d=1.2$ Mbps, $r_u=1.3$ Mbps, $K=100$, and $B=0.5$ kb.

The first aspect we are interested in is identifying the changes in the acquired patients' records at the edge using the proposed patients monitoring scheme. To this end, Figure~\ref{fig:CI} demonstrates the variations in the defined change indicator $\delta$ over different EEG channels for six patients. This figure highlights that using the defined change indicator, a physician can easily interpret the EEG behavior of a patient before, during, and after taking a certain medication. For instance, patients 1, 4, and 5 have a clear increase in their EEG records after taking the medications, while patients 2 and 3 having almost the same behavior before, during, and after taking the medication. Interestingly, our scheme can also detect the errors in collecting the data. For instance, patient 6 has a very large value of $\delta$ for channel 14 only, which indicates that there is a problem in this channel during data collection. Hence, the physician should repeat this experiment for this patient before conducting further data analysis.              

\begin{figure}[b!]
	\centering
		\scalebox{1.83}{\includegraphics[width=0.27 \textwidth]{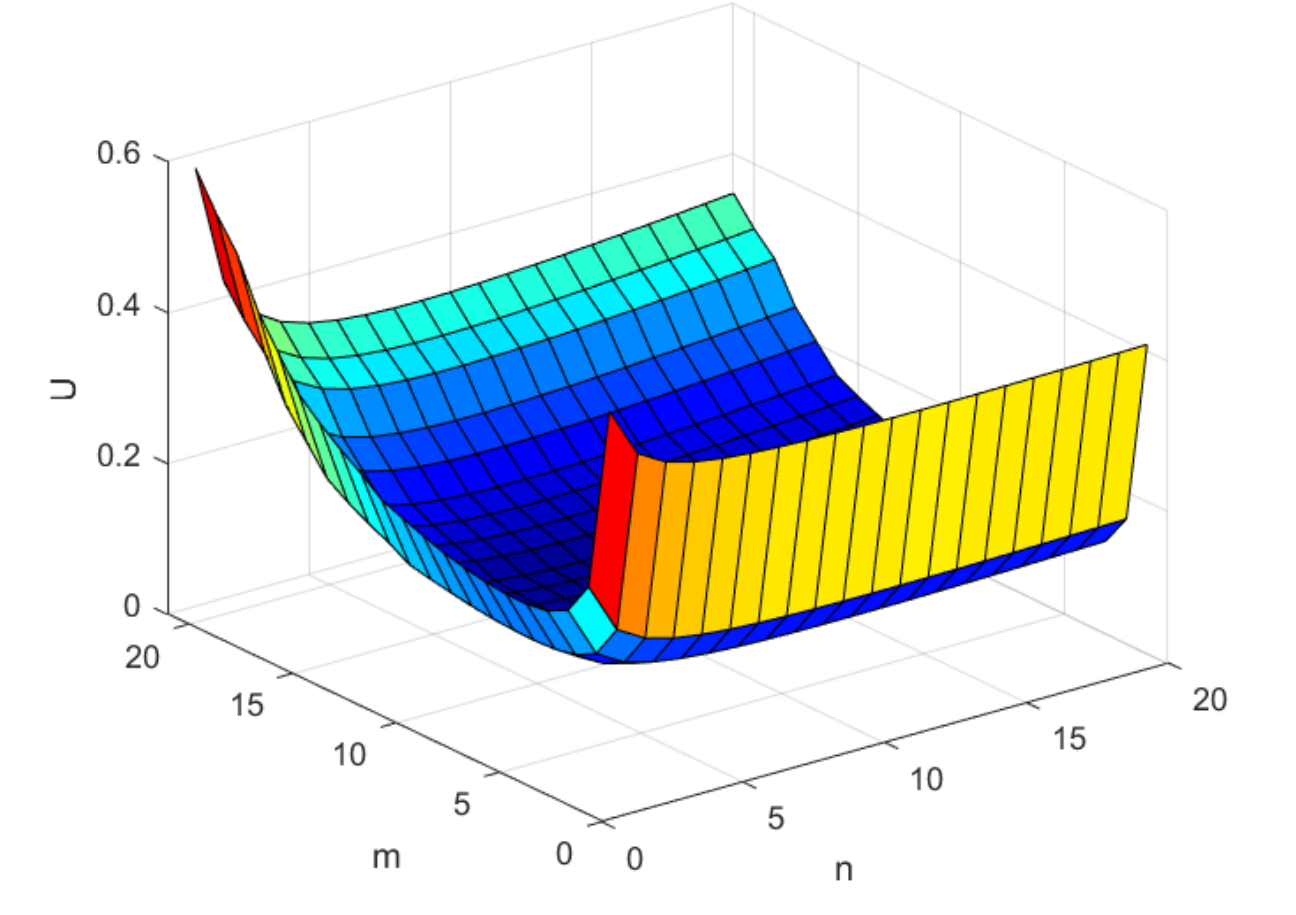}}
	\caption{The proposed objective function as the number of validators ($m$) and the number of transactions per block ($n$) vary, for a one blockchain channel. }
	\label{fig:BlockchainTradeoff}
\end{figure}

\begin{figure}[t!]
	\centering
		\scalebox{1.7}{\includegraphics[width=0.27 \textwidth]{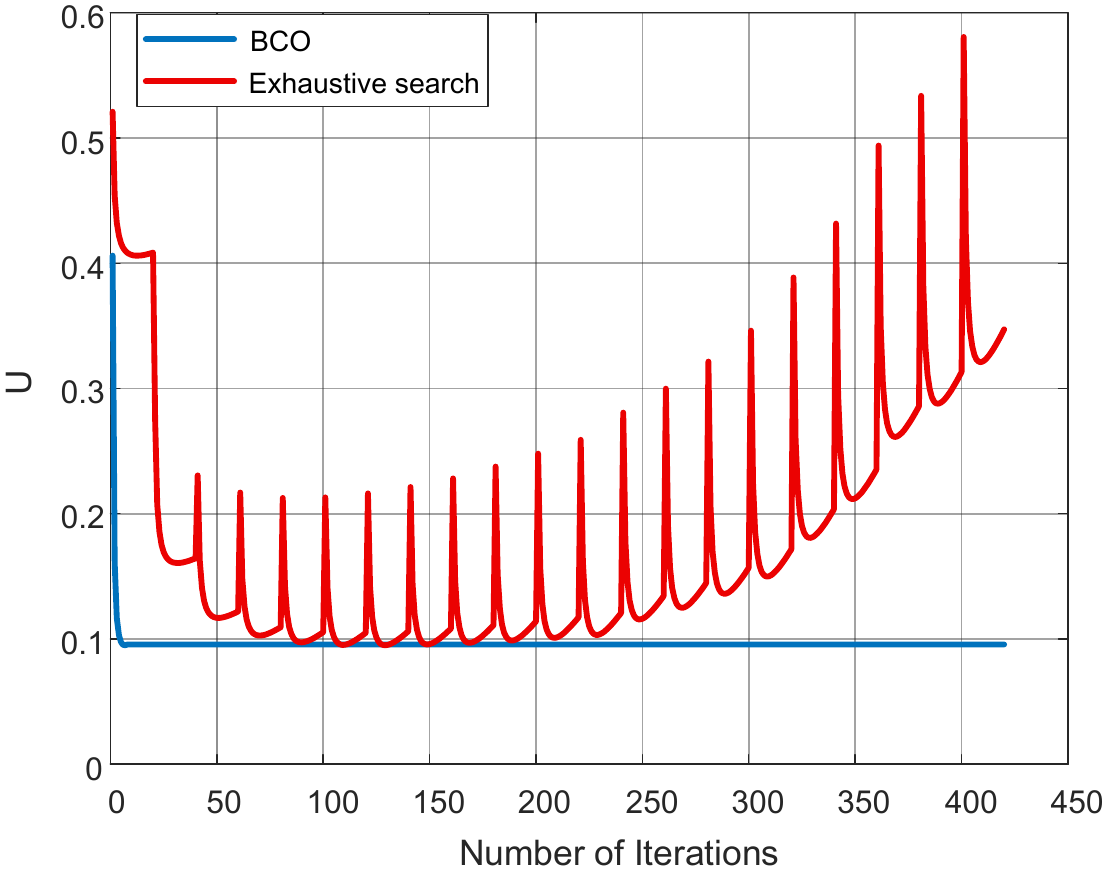}}
	\caption{Convergence behavior of the proposed algorithm compared to the solution obtained through exhaustive search. }
	\label{fig:BlockchainModes}
\end{figure}

\begin{figure*}
	\centering
		\scalebox{3.65}{\includegraphics[width=0.27 \textwidth]{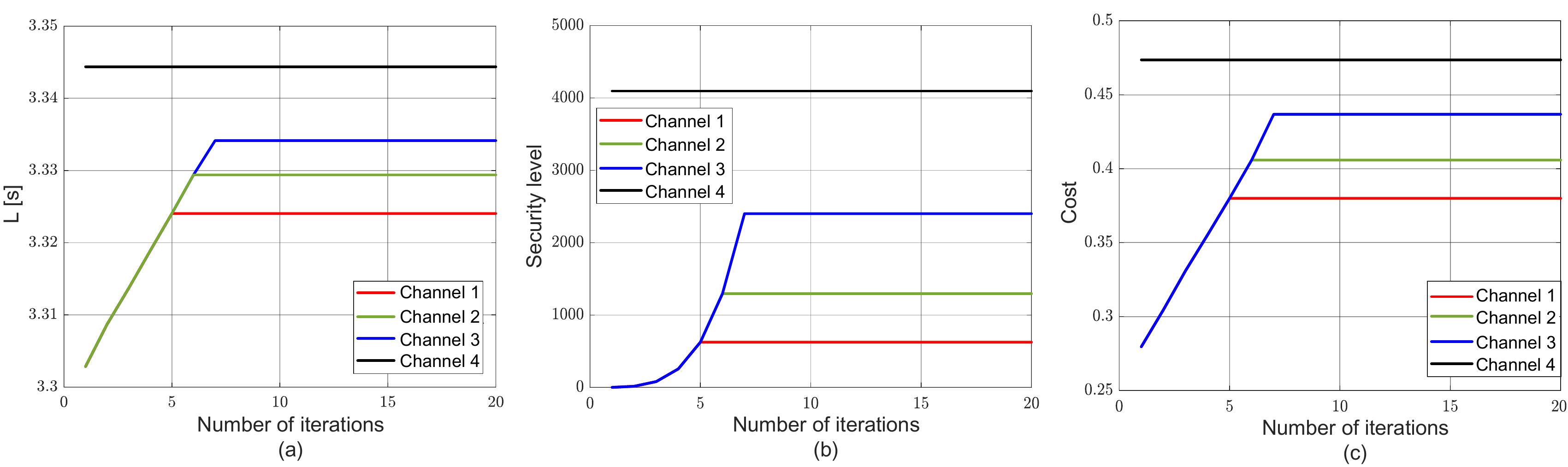}}
	\caption{A comparison of diverse blockchain performance metrics: (a) latency, (b) security, and (c) cost, for various blockchain channels with different configurations. }
	\label{fig:metrics_withchannels2}
\end{figure*}

The second aspect we are interested in is the impact of blockchain configuration optimization on the different performance metrics.  First, Figure~\ref{fig:BlockchainTradeoff} depicts the effect of changing the blockchain configuration parameters (i.e., number of validators $m$ and number of transactions per block $n$) on the obtained utility function in (\ref{eq:utility}), for applications with similar requirements in terms of security, latency, and  cost ($\alpha=\beta=\gamma$). It is clear how changing the configuration parameters always corresponds to a significant change in the utility. Thus, it is important to optimize these parameters considering diverse applications' requirements and system performance.   

As far as the blockchain configuration optimization is concerned, Figure~\ref{fig:BlockchainModes} shows the convergence behavior of the proposed BCO algorithm to the optimal solution obtained by exhaustive search, given $M=21$ and $N=20$. We observe that our algorithm requires only 7 iterations to reach the optimal solution compared to exhaustive search that still does not converge after 420 iterations.   
 
\begin{figure} [t!]
		\scalebox{1.68}{\includegraphics[width=0.27 \textwidth]{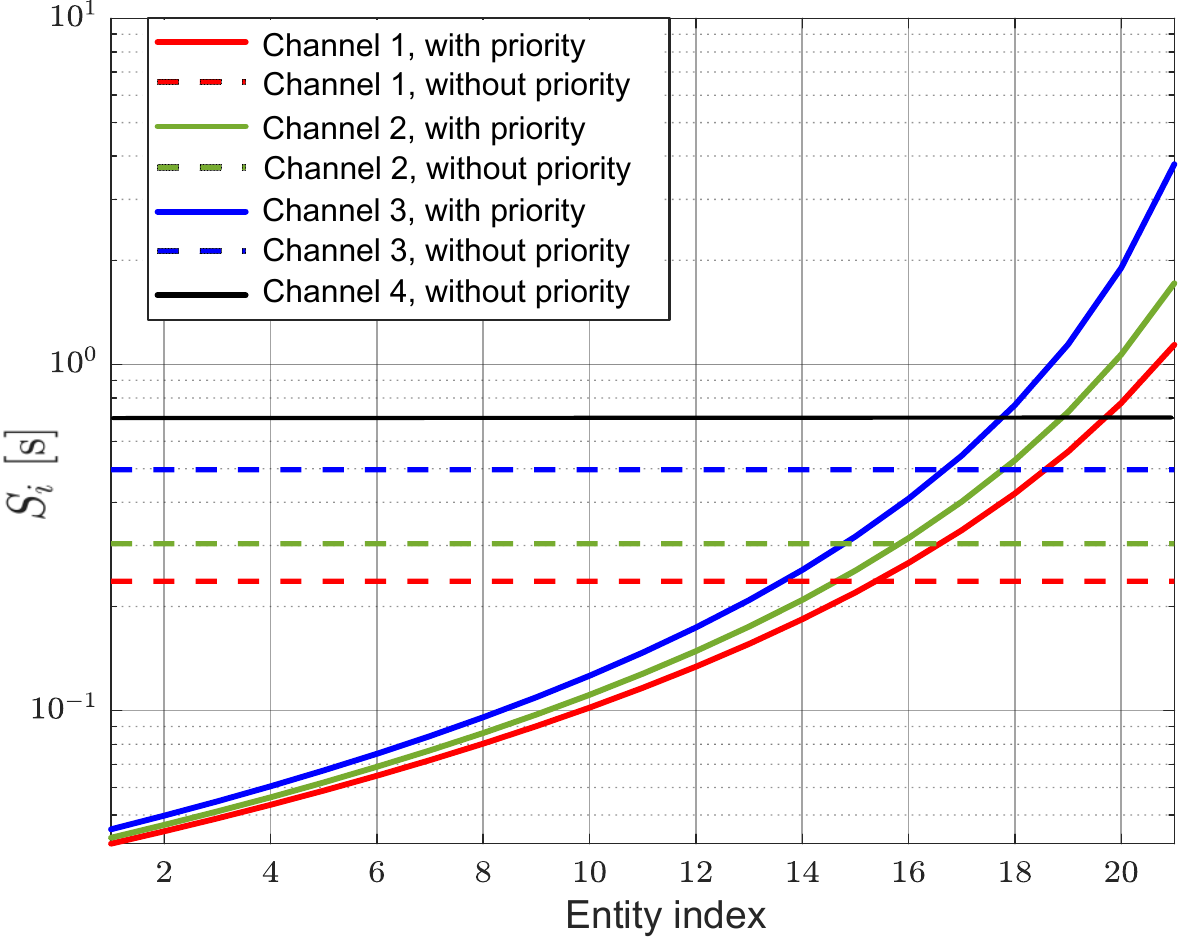}}
	\caption{The obtained average sojourn time for different blockchain channels with and without priority assignments.}
	\label{fig:soujor_withchannels}
\end{figure}

We now study, in Figure~\ref{fig:metrics_withchannels2} and Figure~\ref{fig:soujor_withchannels}, how changing blockchain configuration on different channels influences the performance. The plots in Figure~\ref{fig:metrics_withchannels2} represent the main performance metrics considered in our framework (i.e., latency, security, and cost) as a function of the number of iterations until reaching to the convergence. Each curve therein corresponds to a channel configuration, and each plot corresponds to a performance metric.  
The configuration of the channels from 1 to 3 has been optimized using the proposed BCO scheme, while the configuration of channels 4 is assumed to be fixed, considering a fixed number of validators (i.e., $m=8$) and a fixed number of transactions per block (i.e., $n=80$). Herein, it is assumed that channel 1 is used for urgent data, channel 2 for normal data, and channel 3 for non-urgent data.       
Comparing the individual curves within each plot, we can observe how our BCO algorithm efficiently adjusts different channels configurations according to the acquired data characteristics, such that the urgent data are sent by the lowest latency and computational cost, while the non-urgent data (i.e., require high security without latency constraint) are sent with the highest security level. Moreover, it clearly illustrates the tradeoff between increasing the security level and decreasing the latency. 
Thus, this result shows that it is important to have multiple channels with different configurations within the same blockchain to be able to adapt to diverse types of applications/data with different characteristics. 

Finally, we assess how much, and for whom, our priority assignment scheme is beneficial. Figure~\ref{fig:soujor_withchannels} depicts
how, for different channels configurations, priority assignment influences the obtained sojourn time; different curves correspond to different channels with and without considering priority assignment. This figure highlights that assigning different priorities for different entities in the system (based on the urgency levels or the entity weight) yields a substantial decrease in sojourn time for high-priority entities, hence they can share their transactions with a substantially smaller delay.



\section{Related work and Benefits of I-Health \label{sec:Related}}

This section highlights the key benefits of I-Health, in light of the recent-related literature.

\subsection{Related work \label{sec:Related_work}} 

Outbreak data management have attracted major attention, with several works focusing on monitoring  new virus outbreaks, such as COVID-19 pandemic \cite{Covid_data} and west Africa Ebola epidemic \cite{schafer2016epi}.  
However, large-scale data collection and processing while considering privacy and public trust is challenging \cite{EEGSecurity2019}. Relying on a centralized entity or web resources \cite{websources} for emergency events detection will not be adequate in case of epidemics.   
 Traditionally, public health systems deploy personnel in areas where the epidemic is centered, to collect relevant information. This usually results in physically contacting infected individuals \cite{Epidemiology}. Then, data analysis and epidemic management are performed in a central entity using the received periodic information from the infested areas. For instance, during the severe acute respiratory syndrome (SARS) outbreak in Toronto, an important step to perform seamless outbreak management was building an outbreak management database platform. This platform enables the sharing of public health information, gathering clinical information from hospitals, and integrating them into an interoperable database \cite{SARS2003}. With the help of IoT and recent technologies,  containment and eventual treatment of outbreaks can be run more smoothly. Thanks to the advances of edge computing and blockchain technologies, designing a secure, collaborative health model to implement the integration of multiple national and international entities is now more realizable than ever before.  

The power of security in blockchain comes from the collective resources of the crowd, since, most of the entities have to verify each block of data using a consensus algorithm, e.g. DPoS \cite{AB3}. Hence, any cyber attack has to beat the resources of the whole crowd collectively to be able to hack the integrity of the data, which makes  attacks to the blockchain impractical \cite{PrivacyPreserving, Panacea2018}.    
Recently, different types of blockchain have been envisioned for the healthcare sector, including permissioned and permissionless blockchains. Permissionless blockchains offer decentralized and secure data sharing, however, when advanced control and privacy are required, private or permissioned models turn to be more efficient.  
Several blockchain frameworks (e.g., Ethereum and Hyper ledger Fabric), smart contracts\footnote{A smart contract is a software that contains all instructions and rules agreed upon by all the entities to be applied on the blockchain: all the transactions need to be consistent with the smart contract before being added to the blockchain.}, and consensus algorithms have been investigated in the literature \cite{AB1, AB4, AB5}.   

The blockchain architectures that have been proposed so far in the literature can be broadly classified into two categories: patient-based and entity-based. In patient-based architectures, patients participate in the blockchain \cite{AB2, AB11};  in entity-based architectures, instead, health organizations, hospitals, research institutes, and alike are the main actors, while patients only interact with the health organizations to acquire the service they need \cite{AB12}.  
For instance, \cite{AB3} exploits blockchain to link patients, hospitals, health bureaus, and diverse healthcare communities for enabling comprehensive medical records sharing and review. \cite{AB7} presents a user-centric medical data sharing solution, where a mobile application is used to gather the data from wearable devices, then sharing the data with healthcare providers and insurance companies using permissioned blockchain. \cite{AB9} introduces a blockchain-based system that enables data provenance, auditing, and control over shared medical data between different entities. This system utilizes smart contracts and an access control scheme to detect malicious activities on the shared data and deny access to offending entities.   
However, most of the aforementioned approaches suffer from poor scalability,  computational cost,  and slow response.  
We therefore envision a solution that combines the  blockchain-enabled architecture with intelligent processing at the edge so as to support fast, secure and scalable exchange and processing of medical data. A preliminary version of our study has been presented in \cite{magazine2020}, where only a single-channel blockchain architecture is considered without edge functionality and priority assignment.

\subsection{Benefits of the proposed I-Health system \label{sec:Benefits}} 

In the light of the aforementioned challenges and initiatives, we highlight the practical benefits of leveraging the proposed I-Health system during the epidemics. 

\subsubsection{Infected patients monitoring}
The proposed I-Health system allows for the timely monitoring of the changes in the patients' state and when those changes occur. Leveraging the advances of edge computing and blockchain within I-Health framework enables real-time remote monitoring for quarantined patients. This, in one hand, allows the doctors to communicate with the patients while monitoring their vital signs remotely, and on the other hand, it minimizes the physical interactions between the medical staff and the patients while reducing the patients' flow to the overcrowded hospitals. Moreover, the fast dissemination, processing, and analysis of medical data using I-Health have been perceived to be crucial for speeding up the process of finding adequate medications for emerging diseases. We also highlight that the proposed architecture allows for implementing efficient localization techniques at the edge (such as the one in \cite{Localization}), hence it can enable patients monitoring and tracking, which is important in case of epidemics.     

\subsubsection{Remote accessibility of medical data}  
By supporting a secure, remote access to the patients' EHRs using I-Health, the medical staff can timely review the records from various locations to gather important information about different infected cases. This can significantly accelerate data analysis and health learning curves.	 
Moreover, sharing relevant data between different healthcare entities could help in: providing fast response to epidemics, improving nation wide statistics, and enhancing the quality of service.  

\subsubsection{Patients' flow management}
Optimizing patient flow aims at quickly and effectively fulfilling the demand of healthcare by managing and correlating the data related to the patients across multiple entities. Poorly managed patients flow is not usually due to insufficient resources, but due to inefficient scheduling and resource management. This can be addressed using I-Health, which enables the cooperation between diverse health entities to efficiently allocate the available resources to the forthcoming demands.

\section{Conclusion\label{sec:conclusion}}

Next-generation healthcare systems are being shaped by incorporating emerging technologies to provide radical improvements in healthcare services. Thus, this paper proposes a novel, collaborative I-health system for enabling effective and large-scale epidemics management. The proposed I-Health system leverages IoT, edge computing, and blockchain to provide secure management of large amount of medical data generated by various health entities, while effectively addressing the challenges and requirements posed by epidemics.        
 In particular, we propose an effective method for monitoring the patients, at the edge, to ensure early detection, scalability, and fast response for urgent events.  
Furthermore, we develop an optimized blockchain configuration model with a queuing-based priority assignment scheme to optimally manage the received transactions from diverse entities.  
Our results show that mapping the characteristics of the gathered data onto adequate configurations of the blockchain can significantly improve the performance of the overall I-Health system, while fulfilling different health entities' requirements. 


\balance 

\bibliographystyle{IEEEtrannames}
\bibliography{I_health_For_epidemics}

\begin{thebibliography}{10}
\providecommand{\url}[1]{#1}
\csname url@rmstyle\endcsname
\providecommand{\newblock}{\relax}
\providecommand{\bibinfo}[2]{#2}
\providecommand\BIBentrySTDinterwordspacing{\spaceskip=0pt\relax}
\providecommand\BIBentryALTinterwordstretchfactor{4}
\providecommand\BIBentryALTinterwordspacing{\spaceskip=\fontdimen2\font plus
\BIBentryALTinterwordstretchfactor\fontdimen3\font minus
  \fontdimen4\font\relax}
\providecommand\BIBforeignlanguage[2]{{%
\expandafter\ifx\csname l@#1\endcsname\relax
\typeout{** WARNING: IEEEtran.bst: No hyphenation pattern has been}%
\typeout{** loaded for the language `#1'. Using the pattern for}%
\typeout{** the default language instead.}%
\else
\language=\csname l@#1\endcsname
\fi
#2}}

\bibitem{ransomware}
``Healthcare report for 1st half of 2018,''
  \url{https://www.cryptonitenxt.com/resources}, accessed: 2019-03-05.

\bibitem{WHO}
``Coronavirus disease {(COVID-19)} outbreak,''
  \url{https://www.who.int/emergencies/diseases/novel-coronavirus-2019},
  accessed: 2020-03-21.

\bibitem{WorldEconomicForum}
{}, ``Deep shift: Technology tipping points and societal impact,'' \emph{World
  Economic Forum}, Sep. 2015.

\bibitem{pandamic2007}
Levin and {Peter J et al.}, ``Can the health-care system meet the challenge of
  pandemic flu? planning, ethical, and workforce considerations,'' \emph{Public
  Health Rep.}, vol. 122, no.~5, pp. 573--578, 2007.

\bibitem{Cybersecurity2020}
``{The Cybersecurity 202: Hospitals face a surge of cyberattacks during the
  novel coronavirus pandemic},''
  \url{https://www.washingtonpost.com/news/powerpost/}, accessed: 2020-04-19.

\bibitem{suggested}
S.~{Biswas}, K.~{Sharif}, F.~{Li}, B.~{Nour}, and Y.~{Wang}, ``A scalable
  blockchain framework for secure transactions in {IoT},'' \emph{IEEE Internet
  of Things Journal}, vol.~6, no.~3, pp. 4650--4659, June 2019.

\bibitem{AB3}
S.~Wang, J.~Wang, X.~Wang, T.~Qiu, Y.~Yuan, L.~Ouyang, Y.~Guo, and F.-Y. Wang,
  ``Blockchain-powered parallel healthcare systems based on the {ACP}
  approach,'' \emph{IEEE Transactions on Computational Social Systems},
  vol.~99, pp. 1--9, 2018.

\bibitem{Multichannels}
E.~Androulaki, et~al., ``Hyperledger fabric: a distributed operating system for
  permissioned blockchains,'' in \emph{Proceedings of the Thirteenth EuroSys
  Conference}, 2018, pp. 1--15.

\bibitem{Incentivizing2019}
J.~{Kang}, Z.~{Xiong}, D.~{Niyato}, P.~{Wang}, D.~{Ye}, and D.~I. {Kim},
  ``Incentivizing consensus propagation in proof-of-stake based consortium
  blockchain networks,'' \emph{IEEE Wireless Communications Letters}, vol.~8,
  no.~1, pp. 157--160, 2019.

\bibitem{EEGbook}
T.~Yamada and E.~Meng, \emph{Practical Guide for Clinical Neurophysiologic
  Testing: {EEG}}.\hskip 1em plus 0.5em minus 0.4em\relax Lippincott Williams
  {\&} Wilkins, 2012.

\bibitem{BC2019}
J.~{Kang}, Z.~{Xiong}, D.~{Niyato}, D.~{Ye}, D.~I. {Kim}, and J.~{Zhao},
  ``Toward secure blockchain-enabled internet of vehicles: Optimizing consensus
  management using reputation and contract theory,'' \emph{IEEE Transactions on
  Vehicular Technology}, vol.~68, no.~3, pp. 2906--2920, March 2019.

\bibitem{moreverifier}
C.~{Decker} and R.~{Wattenhofer}, ``Information propagation in the bitcoin
  network,'' in \emph{IEEE P2P 2013 Proceedings}, Sep. 2013, pp. 1--10.

\bibitem{adan2002queueing}
I.~Adan and J.~Resing, \emph{Queueing theory}.\hskip 1em plus 0.5em minus
  0.4em\relax Eindhoven University of Technology, Eindhoven, 2002.

\bibitem{malandrino2019reducing}
F.~Malandrino, C.-F. Chiasserini, G.~Einziger, and G.~Scalosub, ``Reducing
  service deployment cost through vnf sharing,'' \emph{IEEE/ACM Transactions on
  Networking}, vol.~27, no.~6, pp. 2363--2376, 2019.

\bibitem{Cloudprice2019}
Z.~{Xiong}, S.~{Feng}, W.~{Wang}, D.~{Niyato}, P.~{Wang}, and Z.~{Han},
  ``Cloud/fog computing resource management and pricing for blockchain
  networks,'' \emph{IEEE Internet of Things Journal}, vol.~6, no.~3, pp.
  4585--4600, June 2019.

\bibitem{Integeroptimiz}
H.~R., {Köppe M.}, L.~J., and W.~R., \emph{Nonlinear Integer
  Programming}.\hskip 1em plus 0.5em minus 0.4em\relax Springer, Berlin,
  Heidelberg, 2010.

\bibitem{ourdata}
\BIBentryALTinterwordspacing
A.~A. Abdellatif, Z.~Chkirbene, A.~Al-Marridi, A.~Mohamed, A.~Erbad, M.~D.
  O’Connor, J.~Laughton, A.~Villacorte, and J.~Menez, ``{EEG} data for
  patients receiving intravenous antibiotic medication,'' 2020. [Online].
  Available: \url{http://dx.doi.org/10.21227/qcg5-yd65}
\BIBentrySTDinterwordspacing

\bibitem{HMC}
``Hamad medical corporation,'' \url{https://www.hamad.qa/}, accessed:
  2020-04-10.

\bibitem{EPOC}
``{EMOTIV EPOC+},'' \url{https://www.emotiv.com/epoc/}, accessed: 2020-04-10.

\bibitem{Covid_data}
M.~Ienca and E.~Vayena, ``On the responsible use of digital data to tackle the
  {COVID-19} pandemic,'' \emph{Nature Medicine}, March 2020.

\bibitem{schafer2016epi}
I.~J. Schafer, E.~Knudsen, L.~A. McNamara, S.~Agnihotri, P.~E. Rollin, and
  A.~Islam, ``The epi info viral hemorrhagic fever (vhf) application: a
  resource for outbreak data management and contact tracing in the 2014--2016
  west africa ebola epidemic,'' \emph{The Journal of infectious diseases}, vol.
  214, no. suppl\_3, pp. S122--S136, 2016.

\bibitem{EEGSecurity2019}
Y.~{Xiao}, Y.~{Jia}, X.~{Cheng}, J.~{Yu}, Z.~{Liang}, and Z.~{Tian}, ``I can
  see your brain: Investigating home-use electroencephalography system
  security,'' \emph{IEEE Internet of Things Journal}, vol.~6, no.~4, pp.
  6681--6691, 2019.

\bibitem{websources}
Z.~{Xu}, X.~{Luo}, Y.~{Liu}, K.~R. {Choo}, V.~{Sugumaran}, N.~{Yen}, L.~{Mei},
  and C.~{Hu}, ``From latency, through outbreak, to decline: Detecting
  different states of emergency events using web resources,'' \emph{IEEE
  Transactions on Big Data}, vol.~4, no.~2, pp. 245--257, 2018.

\bibitem{Epidemiology}
S.~Rasmussen and R.~Goodman, ``The cdc field epidemiology manual.'' \emph{New
  York: Oxford University Press}, 2019.

\bibitem{SARS2003}
{Public Health Agency of Canada}, ``Learning from {SARS}: Renewal of public
  health in canada,''
  \url{http://www.phac-aspc.gc.ca/publicat/sars-sras/naylor}, accessed:
  2020-03-24.

\bibitem{PrivacyPreserving}
J.~{Xu}, K.~{Xue}, S.~{Li}, H.~{Tian}, J.~{Hong}, P.~{Hong}, and N.~{Yu},
  ``Healthchain: A blockchain-based privacy preserving scheme for large-scale
  health data,'' \emph{IEEE Internet of Things Journal}, vol.~6, no.~5, pp.
  8770--8781, 2019.

\bibitem{Panacea2018}
C.~{Esposito}, A.~{De Santis}, G.~{Tortora}, H.~{Chang}, and K.~R. {Choo},
  ``Blockchain: A panacea for healthcare cloud-based data security and
  privacy?'' \emph{IEEE Cloud Computing}, vol.~5, no.~1, pp. 31--37, 2018.

\bibitem{AB1}
K.~N. Griggs, O.~Ossipova, C.~P. Kohlios, A.~N. Baccarini, E.~A. Howson, and
  T.~Hayajneh, ``Healthcare blockchain system using smart contracts for secure
  automated remote patient monitoring,'' \emph{Journal of medical systems},
  vol.~42, no.~7, 2018.

\bibitem{AB4}
H.~Tian, J.~He, and Y.~Ding, ``Medical data management on blockchain with
  privacy,'' \emph{Journal of medical systems}, vol.~43, no.~2, 2019.

\bibitem{AB5}
A.~Al~Omar, M.~Z.~A. Bhuiyan, A.~Basu, S.~Kiyomoto, and M.~S. Rahman,
  ``Privacy-friendly platform for healthcare data in cloud based on blockchain
  environment,'' \emph{Future Generation Computer Systems}, vol.~95, pp.
  511--521, 2019.

\bibitem{AB2}
L.~Chen, W.~K. Lee, C.~C. Chang, K.~K.~R. Choo, and N.~Zhang, ``Blockchain
  based searchable encryption for electronic health record sharing,''
  \emph{Future Generation Computer Systems}, vol.~95, pp. 420--429, 2019.

\bibitem{AB11}
B.~Shen, J.~Guo, and Y.~Yang, ``{MedChain:} efficient healthcare data sharing
  via blockchain,'' \emph{Applied Sciences}, vol.~9, no.~6, 2019.

\bibitem{AB12}
A.~Zhang and X.~Lin, ``Towards secure and privacy-preserving data sharing in
  e-health systems via consortium blockchain,'' \emph{Journal of medical
  systems}, vol.~42, no.~8, 2018.

\bibitem{AB7}
X.~{Liang}, J.~{Zhao}, S.~{Shetty}, J.~{Liu}, and D.~{Li}, ``Integrating
  blockchain for data sharing and collaboration in mobile healthcare
  applications,'' in \emph{2017 IEEE 28th Annual International Symposium on
  Personal, Indoor, and Mobile Radio Communications (PIMRC)}, Oct 2017, pp.
  1--5.

\bibitem{AB9}
Q.~I. Xia, E.~B. Sifah, K.~O. Asamoah, J.~Gao, X.~Du, and M.~Guizani,
  ``{MeDShare:} trust-less medical data sharing among cloud service providers
  via blockchain,'' \emph{{IEEE} Access}, vol.~5, pp. 14\,757--14\,767, 2017.

\bibitem{magazine2020}
A.~A. Abdellatif, A.~Z. Al-Marridi, A.~Mohamed, A.~Erbad, C.~F. Chiasserini,
  and A.~Refaey, ``{SSHealth}: Toward secure, blockchain-enabled healthcare
  systems,'' \emph{accepted in IEEE Network}, 2020.

\bibitem{Localization}
W.~{Li}, Z.~{Chen}, X.~{Gao}, W.~{Liu}, and J.~{Wang}, ``Multimodel framework
  for indoor localization under mobile edge computing environment,'' \emph{IEEE
  Internet of Things Journal}, vol.~6, no.~3, pp. 4844--4853, 2019.

\end{thebibliography}
\end{document}